\documentclass[conference]{IEEEtran}

\IEEEoverridecommandlockouts                              

\usepackage{cite}
\usepackage{amsmath,amssymb,amsfonts}
\usepackage{algorithmic}
\usepackage{graphicx}
\usepackage{textcomp}
\usepackage{xcolor}
\usepackage{balance}
\usepackage{nicefrac,braket,graphicx,subcaption,xfrac,scalerel,multirow,multicol,qcircuit,makecell,url}
\usepackage[colorlinks = true,
urlcolor  = blue]{hyperref}

\allowdisplaybreaks
\def\BibTeX{{\rm B\kern-.05em{\sc i\kern-.025em b}\kern-.08em
    T\kern-.1667em\lower.7ex\hbox{E}\kern-.125emX}}

\title{Optimization of Sensor-Placement on Vehicles using Quantum-Classical Hybrid Methods}

\author{\IEEEauthorblockN{Sayantan Pramanik$^1$$^*$\thanks{*Corresponding authors.}, Vishnu Vaidya$^1$, Gajendra Malviya$^2$, Sudhir Sinha$^2$, Shripad Salsingikar$^{2*}$, \\ M Girish Chandra$^{2*}$, C V Sridhar$^1$, Godfrey Mathais$^1$, Vidyut Navelkar$^1$}
\IEEEauthorblockA{\textit{$^1$TCS Incubation, $^2$TCS Research,
		TATA Consultancy Services, India}
		\\
		\{sayantan.pramanik, vaidya.vishnu, gajendra.malviya, sudhir.sinha, shripad.salsingikar, m.gchandra, \\ sridhar.cv, godfrey.mathais, vidyut.navelkar\}@tcs.com}
}

\begin{document}

\maketitle

\thispagestyle{empty}
\pagestyle{empty}

\begin{abstract}
	Placement of sensors on vehicles for safety and autonomous capability is a complex optimization problem when considered in the full-blown form, with different constraints. Considering that Quantum Computers are expected to be able to solve certain optimization problems more ``easily" in the future, the problem was posted as part of the BMW Quantum Computing Challenge 2021. In this paper, we have presented two formulations for quantum-enhanced solutions in a systematic manner. In the process, necessary simplifications are invoked to accommodate the current capabilities of Quantum Simulators and Hardware. The presented results and observations from elaborate simulation studies demonstrate the correct functionality and usefulness of the proposals.
\end{abstract}


\section{Introduction}\label{sec:intro}
The paper presents two quantum computing methods to optimize the placement of sensors on the surface of a vehicle. 
The problem, as posted in the BMW Quantum Computing Challenge 2021 \cite{bmw_challenge, bmw_problem}, is to arrive at the optimal configurations (position, type and orientation) of the sensors on the vehicle surface that maximizes coverage of the Region of Interest (RoI), while minimizing the total cost of the selected sensors. 
The dataset contains approximately 100,000 points in the RoI, with defined measures of criticality (ranging between 0 to 1), distributed over a volume of approximately 40,000 cubic metres around the vehicle. 
The types of sensors, their coverage parameters and costs are inputs to the problem. 
The optimization problem is decomposed and solved individually for each side of the vehicle, and the results for the entire vehicle are collated and presented in the paper. 
Further, the constraint of covering the points with criticality greater than 0.7 by at least two sensors is not included in the formulation. 
However, the degree of fulfillment of this constraint is calculated as post-processing and is reported in the results.

The first approach fixes, a priori, the number of sensors that can be selected and uses Variational Quantum Eigensolver (VQE) \cite{vqe} to find the optima. The solution is obtained by iterating over the possible number of sensors and picking the one with the best objective value. In this approach, the number of qubits required grows quasi-logarithmically with the number of configurations. The equivalent Integer Linear Programming (ILP) model is solved using a standard, state-of-the-art ILP solver.

The second approach considers a quadratic approximation to the maximum set-coverage formulation with the advantage of obtaining the optimal number of sensors as an output from the model itself. The resultant convex quadratic program is solved with quantum methods using quantum annealing and VQE, as well as a standard, state-of-the-art Integer Quadratic Program (IQP) solver, respectively. The quadratic approximation consumes qubits that are linear in the number of total configurations and significantly reduces the number of decision variables required in an Ising formulation of the problem.

For both the approaches, quantum and classical methods are found to perform on par with each other. A detailed comparison of the results is presented in the paper. Given the combinatorial nature of the optimization problem, the classical approach may become intractable while solving the problem in an integrated manner (instead of solving individually for each side). Under such scenarios, there is a possibility to expect benefits by adopting the proposed strategies with the availability of large-sized quantum computers in the near future.

The paper is organized as follows: Section \ref{sec:prob} delves deeper into the problem statement and provides details on the data on which the experiments have been conducted, Section \ref{sec:methods} describes the two approaches utilized to solve the problem, along with intermittent results from both, followed by the conclusions in Section \ref{sec:conclusions}.

\section{Problem Statement and Data}\label{sec:prob}
\begin{figure*}[h]
	\centering
	\begin{subfigure}{0.3\textwidth}
		\centering
		\includegraphics[width=\textwidth]{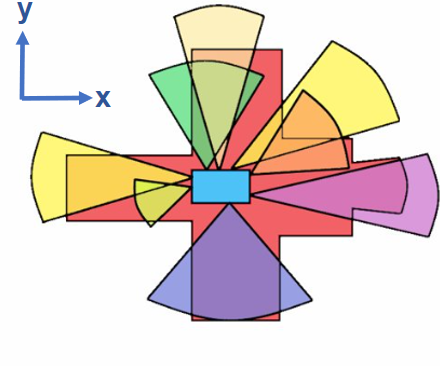}
		\caption{\small{}}
		\label{fig:fov1}
	\end{subfigure}
	\begin{subfigure}{0.3\textwidth}  
		\centering 
		\includegraphics[width=\textwidth]{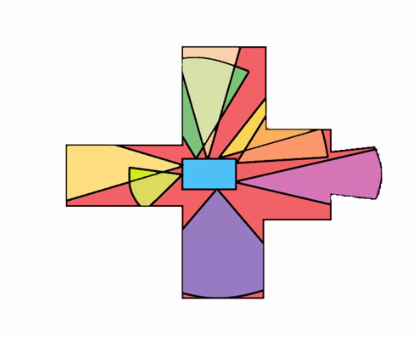}
		\caption{\small{}}
		\label{fig:fov2}
	\end{subfigure}
	\begin{subfigure}{0.3\textwidth}  
		\centering 
		\includegraphics[width=\textwidth]{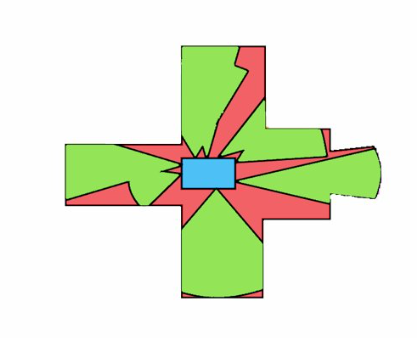}
		\caption{\small{}}
		\label{fig:fov3}
	\end{subfigure}
	\caption[]
	{\small Figure (a) depicts the FoVs of individual sensors placed at different positions of the vehicle (blue rectangle, with \textit{front} towards the positive $x$-direction) at different orientations using various colours. Only the portions of the FoVs that lie within the red RoI adds to the coverage, as shown in (b). The effective-FoV (green) and coverage provided by the sensors is portrayed in (c).} 
	\label{fig:fovs}
	\vspace{-0.5cm}
\end{figure*}
The Sensor Placement problem of BMW Quantum Challenge \cite{bmw_problem}, necessitates the coverage of the space surrounding a vehicle, aptly named as the \textit{Region of Interest} (RoI), through the use of various sensors placed on the vehicle-surface. Different types of sensors utilize various methods to sense the different kinds of data. The aggregated input stream of signals from all the sensors on the vehicle can be converted from unstructured data to structured information and leveraged for making informed (and sometimes autonomous) decisions.

The sensors have parameters, such as the range which signifies how far the sensor can \textit{look} and gauge the environment, as well as its horizontal and vertical angular \textit{sweep}, $\alpha_H$ and $\alpha_V$, respectively (as shown Figure 2 of the challenge statement \cite{bmw_problem}). Further, the sensor can be placed at a particular position and at a given angle with respect to the surface, all of which determine its \textit{Field of View} (FoV), which typically takes the shape of an elliptical cone \cite{vespa}. A point in the RoI is said to be \textit{covered} only if it lies within the FoV of at least one of the selected sensors.

The challenge lies in choosing the appropriate combination of sensor-configurations (type, position and orientation) that can maximally cover the entire RoI. Using too many sensors increases the overall cost and causing an overlap in their FoVs, resulting in a redundancy which ultimately does not improve the coverage \cite{indoor}. The effective-FoV of a number of sensors places on the surface of a vehicle has been portrayed in Figure \ref{fig:fovs}. As a result, the task is to make a trade-off to select the combination of sensor-configurations that maximize the coverage, while minimizing their cost, simultaneously.

The challenge dataset \cite{bmw_data} provides, as RoI, the coordinates of about a $100,000$ points spread over a volume of $40,000$ cubic metres in the vicinity of the vehicle. The points are discretized at separations of $0.5$ metres, and each point has a \textit{criticality index} which determines how \textit{important} covering that point is. The continuous-valued criticality index ranges from $0$ to $1$, with $0$ having the lowest importance and $1$ carrying the highest. The problem statement also inspires the use of four different types of sensors - Camera, LiDAR, Radar and Ultrasonic. The parameters in Table \ref{tab:sensor_params} were collated from various sources \cite{sensor1, sensor2, sensor3}, while avoiding situations where a sensor type can completely \textit{dominate} another in terms of the cost and the coverage provided, and utilized during experimentation:
\begin{table}[ht]
	\renewcommand\arraystretch{1.5}
	\centering
	\resizebox{0.48\textwidth}{!}{
		\begin{tabular}{| c | c | c | c | c |}
			\hline
			\textbf{SensorType } & $\boldsymbol{\alpha_H(^\circ)}$  & $\boldsymbol{\alpha_V(^\circ)}$ & \textbf{Range $\boldsymbol{(m)}$} & \textbf{Cost (\$)} \\
			\hline
			\textbf{LiDAR} & 80 & 40 &120 & 200 \\ 
			\hline
			\textbf{Radar} & 60 & 5 & 120 & 100 \\
			\hline
			\textbf{Camera} & 90 & 60 & 20 & 120 \\
			\hline
			\textbf{Ultrasonic} & 90 & 5 & 10 & 20 \\
			\hline
		\end{tabular}
	}
	\caption{\small{Parameters considered for the various types of sensors.}}
	\label{tab:sensor_params}
	\vspace{-0.25cm}
\end{table}

From a top-view, $2$D visualization of the RoI (see Figure 4 of the challenge document \cite{bmw_problem}), it is clear that the points in regions pertaining to the four sides - front, back, left and right - of the vehicle are mutually exclusive, with only a minor overlap between the front and the left sides. Exploiting this observation, the problem was decomposed into finding the optimal configurations of the sensors independently for each side, before coalescing the results from all sides to check for the coverage and cost for the overall vehicle. For simplicity, the orientation of the sensors may be fixed such that they always face perpendicularly away from the surface, and results for both - with and without this simplification - have been included in the paper. Finally, the challenge statement also had a fleeting mention about having to cover the points with criticality greater than $0.7$ with at least two sensors of different types. This constraint was disregarded in the paper as it could make the optimization problem significantly more difficult. However, although this constraint was not considered during model-formulation, the extent of fulfilment of this requirement with the said models has also been reported in  Section \ref{sec:methods}.

\section{Methodology and Results}\label{sec:methods}
In this section, we describe the two approaches used to solve the given problem in detail, along with intermittent results and comparisons between the performance of the quantum against the classical models.

\subsection{Fixed Sensor-Count based Formulation}\label{sec:method1}
The problem described in Section \ref{sec:prob}, for each side of the vehicle, was expressed as an Integer Linear Problem (ILP) which aims to minimize an objective function composed of a linear combination of the coverage $V_{cov}$ provided by, and the cost $C$ of selected configurations of sensors which is given by:
\begin{equation}\label{eq:obj1}
\begin{aligned}
\min_{x_{t,p,o}} J &= -\lambda_1 \cdot V_{cov} + \lambda_2 \cdot C \\
&= -\lambda_1 \frac{\sum_r c_r z_r}{\sum_r c_r} + \lambda_2 \sum_{t,p,o}\mathcal{C}_{t} x_{t,p,o} 
\end{aligned}
\end{equation}
where, $T$, $P$ and $O$ are the sets of the possible sensor types, positions and orientations respectively, $t \in T$, $p \in P$, and $o \in O$ characterize the configuration of a sensor; 
The set $P$ is derived as each point of a $G_1 \times G_2$ grid, mapped on each side of the vehicle. 
This descritization reduced the problem search space.
$c_r$ is the criticality index of a point $r$ in the RoI, $z_r \in \{0,1\}$ indicates whether $r$ is covered by at least one sensor or not. 
$\mathcal{C}_{t}$ is the cost of a sensor of the type $t$, and $x_{t,p,o} \in \{0,1\}$ is the binary decision variable to signify whether or not a sensor of configuration $(t,p,o)$ is selected. 
$\lambda_1$ and $\lambda_2$ are the weight factors for the coverage and cost terms, respectively. 
The above is subjected to the following constraints:
\begin{equation}\label{eq:pos_cons}
\sum_{t,o}x_{t,p,o} \leq 1, \; \forall p 
\end{equation}
i.e., at most one sensor can be placed at the position $p$, and,
\begin{equation}
z_r \leq \sum_{t,p,o} x_{t,p,o}y^{t,p,o}_r, \; \forall r
\end{equation}
where, $y^{t,p,o}_r$ is an indicator which denotes whether or not the sensor with configuration $(t,p,o)$ covers the point $r$. This constraint forces the value of $z_r$ to $0$ if $r$ is not covered by any of the selected sensors, while the presence of $z_r$ in the objective function makes $z_r=1$ more favourable, thereby optimizing the value of $x_{t,p,o}$. Additionally, to make the quantum model (described later) simpler, the number of sensors that can be placed on a side of the vehicle was fixed, a priori, to a number $n_s$, such that,
\begin{equation}\label{eq:n_s_cons}
\sum_{t,p,o}x_{t,p,o} = n_s
\end{equation}
The FoVs for all the possible sensor configurations was precomputed \cite{itsc}, and for the selected combinations of configurations at each step of the optimization process, the overlap amongst the FoVs and with the RoI was calculated to evaluate the objective function.

The quantum approach, which employs variational quantum eigensolver (VQE) \cite{vqe} to solve the problem, relies broadly on the same ILP model, with minor differences in the way the constraints are enforced and the optimization procedure is carried out. The process starts with initializing $\lceil \log_2G_1 \rceil + \lceil \log_2G_2 \rceil + \lceil \log_2|T| \rceil + \lceil \log_2|O| \rceil$ qubits in the quantum circuit to the equal, zero-phase superposition state. The number of qubits consumed by the model grows approximately logarithmically with an increase in possible configurations. After the application of the ansatz parametrized by $\boldsymbol{\theta}$, the circuit is executed a number of times and the qubits are measured, resulting in a histogram over all the possible sensor-configurations. The set $S(\boldsymbol{\theta})$ of configurations with the highest probabilities are selected from the distribution, such that the constraints in Equations \eqref{eq:pos_cons} and \eqref{eq:n_s_cons} are satisfied, and $|S(\boldsymbol{\theta})| = n_s$. Alternatively, similar results can be achieved by drawing $n_s$ feasible samples from the quantum circuit. The objective function $J(\boldsymbol{\theta})$, corresponding to $S(\boldsymbol{\theta})$, is calculated classically using Equation \eqref{eq:obj1}. The indirect use of the qubit-measurements to obtain the objective value makes it difficult to adopt the objective function w.r.t the optimizable parameters in the ansatz. The gradient-free, classical optimizer COBYLA was thus used for the parameter-tuning process. The routine is repeated until convergence, at which point, the optimized parameters $\boldsymbol{\theta}^*$ may be utilized to draw the ideal combination of sensors $S(\boldsymbol{\theta}^*)$ from the circuit.

\begin{figure}[h]
	\[ \Qcircuit @C=2em @R=1em @!R{
		\lstick{} & \gate{R_Y(\theta_0)} &	\ctrl{1} &	\qw & 		\qw &		\targ &		\qw\\
		\lstick{} & \gate{R_Y(\theta_1)} &	\targ &		\ctrl{1} & 	\qw &		\qw &		\qw\\
		\lstick{} & \gate{R_Y(\theta_2)} &	\qw &		\targ & 	\ctrl{1} &	\qw &		\qw\\
		\lstick{} & \gate{R_Y(\theta_3)} &	\qw &		\qw &		\targ &		\ctrl{-3} &	\qw \gategroup{1}{2}{4}{2}{.7em}{--}
	} \]
	\caption{\small{First layer of the Basic Entangling Layers ansatz acting on $4$ qubits. The parametrized rotation gates are followed by multiple \textit{CNOT} gates, the target qubit of each gate is determined using the relation $t = (c+l+1) \; mod \; n$, where $c$ and $t$ are the control and target qubit numbers, $l$ is the ansatz-layer being constructed and $n$ is the number of qubits the ansatz operates on. An $L$-layer ansatz avails $nL$ optimizable parameters.}}
	\label{fig:basic}
	\vspace{-0.4cm}
\end{figure}
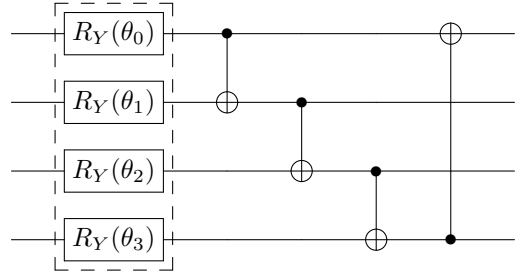

During experimentation, the values of $\lambda_1$ and $\lambda_2$ were fixed at $1$ and $10^{-4}$ to emphasize coverage (which ranges between $0$ and $1$) over the cost (which is typically in the range of hundreds of dollars), respectively, and $4 \times 4$ grids on each side of the vehicle were considered where the sensors are allowed to be placed at a single, fixed orientation, only perpendicular to the surface. For this chosen grid size, $n_s$ was varied from $1$ to $8$ and independent simulations were carried out for each. 
The IBM\textsuperscript{\tiny\textregistered} ILOG\textsuperscript{\tiny\textregistered} CPLEX\textsuperscript{\tiny\textregistered} solver was utilized to solve the ILP.
Qiskit \cite{qiskit} was used to create the circuits with three layers of the basic entangling layers ansatz, as shown in Figure \ref{fig:basic}, which were executed for $1000$ shots on the local QASM simulator. 
Due to the inherently stochastic nature of quantum, all the quantum experiments were run $10$ times. 
The worst outlier was removed, and the average of the rest along with the min-max range are reported in all of the subsequent plots. 
Further, it was noticed that the objective values of the results from the best quantum experiments almost always coincide with the classical ones. 

\begin{figure}
	\begin{subfigure}{0.24\textwidth}
		\hspace{-0.3cm}
		\includegraphics[width=0.19\textheight]{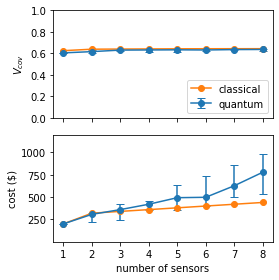}
		\caption{\small{}}
		\label{fig:left_tp}
	\end{subfigure}
	\begin{subfigure}{0.24\textwidth}
		\includegraphics[width=0.19\textheight]{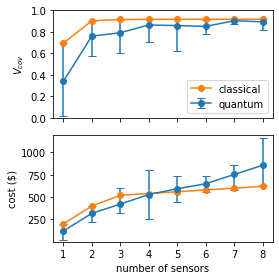}
		\caption{\small{}}
		\label{fig:left_tpo}
	\end{subfigure}
	\caption[ The average and standard deviation of critical parameters ]
	{\small{Comparison of coverage and cost of selected sensor configurations obtained form the classical and quantum solutions for the left side (discretized into $4 \times 4$ grids) of the vehicle, when (a) the orientation of the sensors was fixed perpendicular to the surface, (b) the sensors could be oriented at four different angles.}}
	\label{fig:results}
	\vspace{-0.6cm}
\end{figure}

The plots in Figure \ref{fig:left_tp} show the values of coverage and cost from the quantum and classical experiments for various values of $n_s$ on the $x$-axis for the left side of the vehicle, which are fairly representative of the trend of outcomes observed on the other sides as well. 
The coverage provided by, and the cost of the combination of sensors with the best objective value have been summarized in Table \ref{tab:model1_tp}. Having found the most appropriate combination from each of the sides, the aggregated coverage for the entire RoI and cost of all the selected sensors for both the classical and quantum approaches are captured. In Table \ref{tab:model1_tp}, it is interesting to observe that the aggregate coverage provided by the selected sensors is somewhat higher than a criticality-weighted average of the coverages from the four sides. This can be attributed to the overlap of the RoIs of the sub-problems. For instance, although a sensor placed on the left side may not cover a significant portion of the left-RoI that intersects with the front, another sensor placed at the front may well extend coverage to the region in question.
\begin{table}[ht]
	\renewcommand\arraystretch{1.5}
	\centering
	\resizebox{0.48\textwidth}{!}{
		\begin{tabular}{| c | c | c | c | c |}
			\hline
			\multirow{2}{*}{\textbf{Side}} & \multicolumn{2}{| c |}{\textbf{Classical}} & \multicolumn{2}{| c |}{\textbf{Universal Quantum}} \\
			\cline{2-1}\cline{3-1}\cline{4-1}\cline{5-1}
			& \textbf{Coverage \%} & \textbf{Cost (\$)} & \textbf{Coverage \%} & \textbf{Cost (\$)} \\
			\hline
			\textbf{Front} & 84.60 & 320 & 84.09 & 320 \\ 
			\hline
			\textbf{Left} & 63.84 & 320 & 63.81 & 340 \\
			\hline
			\textbf{Right} & 72.98 & 380 & 71.39 & 440 \\
			\hline
			\textbf{Back} & 76.90 & 380 & 75.62 & 340 \\
			\hline
			\textbf{Aggregate} & \textbf{89.53} & \textbf{1400} & \textbf{88.34} & \textbf{1440} \\
			\hline
		\end{tabular}
	}
	\caption{\small{Comparison of coverage and cost obtained from classical and quantum models, for the combination of sensors with the best objective value with $4 \times 4$ grids, and fixed, perpendicular orientation.}}
	\label{tab:model1_tp}
	\vspace{-0.25cm}
\end{table}

\begin{figure}
	\begin{subfigure}{0.24\textwidth}
		\hspace{-0.3cm}
		\includegraphics[width=0.19\textheight]{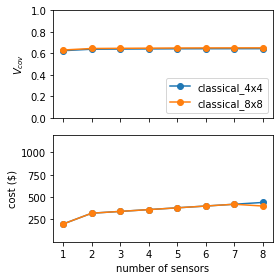}
		\caption{\small{}}
		\label{fig:classical_4vs8}
	\end{subfigure}
	\begin{subfigure}{0.24\textwidth}
		\includegraphics[width=0.19\textheight]{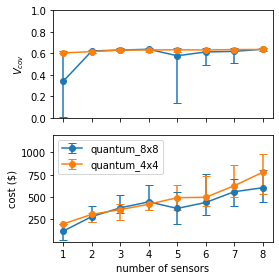}
		\caption{\small{}}
		\label{fig:quantum_4vs8}
	\end{subfigure}
	\caption[ The average and standard deviation of critical parameters ]
	{\small{Comparison of coverage and cost for $4 \times 4$ and $8 \times 8$ discretization on the left side, with all sensors oriented perpendicularly outwards to the surface, for (a) classical and (b) quantum method.\newline}}
	\label{fig:4vs8}
	\vspace{-0.6cm}
\end{figure}

The advantages obtained by opting for a higher discretization of $8 \times 8$ grids on each side of the vehicle were miniscule, as shown in Figure \ref{fig:4vs8} for the left side. There is much to gain, however, by allowing the sensors to be placed at different angles of orientation. This is apparent in the given data on the left and right sides, where the RoI is not symmetric with respect to the vehicle (Figure 4 in \cite{bmw_problem}). With the subscripts $\mathcal{F}$, $\mathcal{B}$, $\mathcal{L}$ and $\mathcal{R}$ denoting the front, back, left and right sides of the vehicle, respectively, the various angles that were considered are as follows: $O_\mathcal{F}=\{-30^\circ , 0^\circ , 30^\circ, 45^\circ \}$, $O_\mathcal{B}=\{30^\circ , 0^\circ , -30^\circ, -45^\circ \}$, $O_\mathcal{L}=\{-60^\circ , -40^\circ , -20^\circ, 0^\circ \}$ and $O_\mathcal{R}=\{0^\circ , 20^\circ , 40^\circ, 60^\circ \}$, which were chosen to facilitate a higher overlap of the FoVs with the RoI. Here, all the angles are in degrees, about an axis parallel to the $z$-direction (as shown in Figure \ref{fig:fov1}). The normal to the surface is taken as the reference, and clockwise angles (as observed from the top-view of the vehicle) are considered as negative. With these values as the possible angles of orientation, and for $4 \times 4$ grid on each side, the results are tabulated in Table \ref{tab:model1_tpo} and the plot for the left side is illustrated in Figure \ref{fig:left_tpo}. The improvement in coverage for each side (for a slight increase in the cost) provided by the free-orientation is significant. Although, similar to the fixed-orientation case, the aggregated-coverage here is ``greater than the sum of the parts", the difference is not as pronounced.

\begin{table}[ht]
	\renewcommand\arraystretch{1.5}
	\centering
	\resizebox{0.48\textwidth}{!}{
		\begin{tabular}{| c | c | c | c | c |}
			\hline
			\multirow{2}{*}{\textbf{Side}} & \multicolumn{2}{| c |}{\textbf{Classical}} & \multicolumn{2}{| c |}{\textbf{Universal Quantum}} \\
			\cline{2-1}\cline{3-1}\cline{4-1}\cline{5-1}
			& \textbf{Coverage \%} & \textbf{Cost (\$)} & \textbf{Coverage \%} & \textbf{Cost (\$)} \\
			\hline
			\textbf{Front} & 89.07 & 440 & 85.98 & 440 \\ 
			\hline
			\textbf{Left} & 90.33 & 400 &90.30 & 600 \\
			\hline
			\textbf{Right} & 96.28 & 520 & 94.80 & 620 \\
			\hline
			\textbf{Back} & 80.52 & 360 & 79.29 & 460 \\
			\hline
			\textbf{Aggregate} & \textbf{92.60} & \textbf{1720} & \textbf{91.60} & \textbf{2120} \\
			\hline
		\end{tabular}
	}
	\caption{\small{{Comparison of coverage and cost obtained from classical and quantum models, for the combination of sensors with the best objective value with $4 \times 4$ grids, and $4$ different orientations.}}}
	\label{tab:model1_tpo}
	\vspace{-0.5cm}
\end{table}

\subsection{Approximate Set-Coverage based Formulation}
In this section we present the maximum set-coverage based quantum approach to solve the sensor placement optimization problem. A conventional Ising formulation of the max set-coverage problem is qubit-intensive.
It's approximate version presented here is expressed as a convex quadratic program with linear constraints, and is more resource friendly, making it conducive to the present Noisy Intermediate-Scale Quantum (NISQ) era \cite{NISQ}. 

A version of the closely-related set cover problem where the ``universe" set $U$ has $n$ elements, $N$ subsets are available to choose from, and $M \leq N$ sets are finally chosen, would lead to $N+n\lceil \log_2 M \rceil$ number of binary decision variables to formulate it as a Quadratic Unconstrained Binary Optimization (QUBO) problem \cite{ising_formulations}. Here, the universe set is analogous to the set of points in the RoI that need to be covered by the sensors, and the given subsets may be equated with the FoV of the sensors arranged in all possible configurations. The task is to find the subset of sensors that reasonably covers the RoI, while also minimizing the monetary cost of the selected set of sensors.

The number of qubits required to solve the set-cover version of the problem increases linearly with the cardinal number of the universe or RoI set, which typically ranges into tens or hundreds of thousands in the real-world. 
Solving the set-cover problem in its entirety may not be possible using quantum computing until there is a drastic increase in the count of qubits available. 
Hence, we resort to the following approximation, through the use of classical preprocessing, to reduce the qubit-requirements to the range of possible sensor configurations:

Let $R=\{r \, | \, r \textrm{ is a point in the RoI}\}$, $F_i$ be the set of points covered by the FoV of a sensor with configuration (type, position and orientation) $i \in \mathcal{S}$, the set of all possible configurations, and $|\mathcal{S}|=N$. Unlike the traditional maximum coverage problem where the number of subsets that can be chosen is bounded by a predefined number $k$, we let it become a free parameter in our formulation, allowing it to be decided during the optimization process. Also, let us assume that all points in the RoI have equal, unit criticality, an assumption that will be generalized later.

The coverage obtained through a collection of sensors can then be expressed as:
\begin{equation}
\label{eq:vcov_full}
V_{cov} = \left| R \cap \left(\bigcup\limits_{i=1}^{N} F_i \circ x_i \right) \right| 
\end{equation}
where $|\mathbb{S}|$ represents the cardinality of any random set $\mathbb{S}$, $x_i$ is a binary variable which denotes whether or not the sensor with configuration $i$ is selected, 
\begin{equation}
F_i \circ x_i = 
\begin{cases}
F_i,& \text{if } x_i=1\\
\O ,& \text{if } x_i=0
\end{cases}
\end{equation}
and
\begin{equation}
|F_i \circ x_i| = |F_i|x_i
\end{equation}

Now, the cardinality of the union of $N$ sets can be approximated through the use of Bonferroni's inequality \cite{bonferroni} with only single and pairwise terms as:
\begin{equation}\label{eq:bon}
\left|  \bigcup\limits_{i=1}^{N} F_i \right|   \geq \sum_i \left| F_i\right|  - \sum_{i<j}\left| F_i \cap F_j\right| 
\end{equation}
which provides a lower bound to the actual coverage accorded by the selection of sensors. Since only pairwise terms are used, a large intersection of FoVs of three or more sensors is indirectly discouraged by the formulation. Further, the maximization of $V_{cov}$ occurs through the maximization of its lower bound. By exploiting the relation in Equation \eqref{eq:bon}, and replacing the cardinality $|\mathbb{S}|$ to the criticality-weighted cardinality $|\mathbb{S}|_w$, Equation \eqref{eq:vcov_full} is simplified to:
\begin{equation}
\label{eq:vcov}
v_{cov} = \sum_i \left| R \cap F_i \circ x_i\right|_w  - \sum_{i<j} \left| R \cap F_i \circ x_i \cap F_j \circ x_j\right|_w
\end{equation}
where $V_{cov} \geq v_{cov}$, and if $\mathbb{S}=\{s_1,\dots,s_m\}$ is a subset of $R$, then,
\begin{equation}
|\mathbb{S}|_w = \frac{\sum_{j=1}^{m} c_{s_j}}{\sum_r c_r}
\end{equation}
The terms $\mu_i = \left| R \cap F_i\right|_w$ and $\sigma_{ij} = \left| R \cap F_i \cap F_j\right|_w$ are pre-computed for every single and pairwise configuration of sensors. As the possible configurations are typically limited by the spatial resolution, types of sensors and their orientations, such pre-calculations can be comfortably be handled by classical computers even for large RoIs. The number of binary decision variables is correspondingly reduced from $N+n\lceil \log_2 M \rceil$ to $N$. As typically $n \gg N$, this makes the proposed approximate set-coverage formulation quite resource-friendly. The expression for coverage is thus approximated by the following quadratic form:
\begin{equation}
\begin{gathered}
v_{cov} = \frac{3}{2} \boldsymbol{\mu}^T\boldsymbol{x} - \frac{1}{2}\boldsymbol{x}^T \boldsymbol\sigma\boldsymbol{x} \\
\boldsymbol{x} \in \{0,1\}^N 
\end{gathered}
\end{equation}
The factor of $\nicefrac{1}{2}$ before the quadratic term is to account for the symmetric nature of $\boldsymbol{\sigma}$, and the coefficient of $\nicefrac{3}{2}$ takes care of the $\mu_i = \sigma_{ii}$ elements in the diagonal of the $\boldsymbol{\sigma}$ matrix. It is trivial to notice that the similarities between $\boldsymbol\sigma$ and covariance matrices, considering that the (weighted) cardinality of intersection of sets can be expressed as $\boldsymbol{X}^T\boldsymbol{X}$, where the $n$-length columns of $\boldsymbol{X}$ contain the square-roots of the criticality indices for the points covered by the FoVs, and zero for the points that are left uncovered. As $\boldsymbol{\sigma}$ is symmetric and positive-semidefinite, the defined $v_{cov}$ has a convex quadratic form. Further, with the above expression of $\boldsymbol{\sigma}$ as $\boldsymbol{X}^T\boldsymbol{X}$, the entries may even be calculated using quantum techniques which provide a benefit over classical inner product calculation \cite{inner_product}, given an efficient way to load the data.

The sensor configuration $i$ can be resolved into the sensor's type ($t$), position ($p$), and orientation ($o$), and all variables assume binary values, making the resultant Integer Quadratic Program (IQP) NP-hard \cite{np-hard}:
\begin{equation}\label{eq:model2}
\begin{aligned}
\min_{x_{t,p,o}} \quad & -\lambda_1 v_{cov}  + \lambda_2 \sum_{t,p,o}\mathcal{C}_{t} x_{t,p,o} \\
\textrm{s.t.} \quad & \sum_{t,o} x_{t,p,o} \leq 1, \; \forall p \\
& x_{t,p,o} \in \{0,1\}
\end{aligned}
\end{equation}
To find the minima of the above convex quadratic program, classical, quantum annealing, and gate-model based methods were employed. As before, we stick to the weights $\lambda_1=1$ and $\lambda_2=10^{-4}$, across all the approaches. 
Then IBM\textsuperscript{\tiny\textregistered} ILOG\textsuperscript{\tiny\textregistered} CPLEX\textsuperscript{\tiny\textregistered} was used to solve the IQP.
For experimentation on quantum annealing simulators, the objective function and the constraints in Equation \eqref{eq:model2} were converted to Binary Quadratic Model (BQM) and $1000$ samples were drawn using the simulated annealing sampler on D-Wave Ocean SDK \cite{dwave}. 

For both, the classical and annealer-based approaches, experiments were separately conducted both with and without considering orientation as a decision variable, as done before. The slack variables, that are introduced during the inclusion of the inequality constraint into the QUBO formulation, add an overhead to the qubit-count. Amidst the already-scarce resources, to keep the qubit requirements in check, the performance of the models with and without considering the constraint was compared, and both seemed to return feasible and almost similar results. Further, each of the $4 \times 4$ grids on the vehicle's sides are large enough to accommodate multiple sensors, even if the constraint in Equation \ref{eq:pos_cons} was violated. Keeping these points in mind, the constraint was foregone for both the quantum and classical models.
The discretization and orientation-angles discussed in Section \ref{sec:method1} were reused, and results from the classical and annealer-based models for the fixed-orientation ($t,p$) and free-orientation ($t,p,o$) cases are presented in Table \ref{tab:results_2_tp} and Table \ref{tab:results_2_tpo}, respectively. 

\begin{table}[ht]
	\renewcommand\arraystretch{1.5}
	\centering
	\resizebox{0.48\textwidth}{!}{
		\begin{tabular}{| c | c | c | c | c |}
			\hline
			\multirow{2}{*}{\textbf{Side}} & \multicolumn{2}{| c |}{\textbf{Classical}} & \multicolumn{2}{| c |}{\textbf{Quantum Annealer}} \\
			\cline{2-1}\cline{3-1}\cline{4-1}\cline{5-1}
			& \textbf{Coverage \%} & \textbf{Cost (\$)} & \textbf{Coverage \%} & \textbf{Cost (\$)} \\
			\hline
			\textbf{Front} & 84.60 & 320 & 84.60 & 320 \\ 
			\hline
			\textbf{Left} & 63.84 & 320 & 63.84 & 320 \\
			\hline
			\textbf{Right} & 71.81 & 320 & 71.81 & 320 \\
			\hline
			\textbf{Back} & 75.40 & 260 & 75.39 & 260 \\
			\hline
			\textbf{Aggregate} & \textbf{89.21} & \textbf{1220} & \textbf{88.85} & \textbf{1220} \\
			\hline
		\end{tabular}
	}
	\caption{\small{{Results from the classical and annealer-based approximate max set-coverage models with $4 \times 4$ grids, for the fixed-orientation case.}}}
	\label{tab:results_2_tp}
	\vspace{-0.5cm}
\end{table}

\begin{table}[ht]
	\renewcommand\arraystretch{1.5}
	\centering
	\resizebox{0.48\textwidth}{!}{
		\begin{tabular}{| c | c | c | c | c |}
			\hline
			\multirow{2}{*}{\textbf{Side}} & \multicolumn{2}{| c |}{\textbf{Classical}} & \multicolumn{2}{| c |}{\textbf{Quantum Annealer}} \\
			\cline{2-1}\cline{3-1}\cline{4-1}\cline{5-1}
			& \textbf{Coverage \%} & \textbf{Cost (\$)} & \textbf{Coverage \%} & \textbf{Cost (\$)} \\
			\hline
			\textbf{Front} & 86.37 & 360 & 86.67 & 380 \\ 
			\hline
			\textbf{Left} & 90.33 & 400 & 90.30 & 400 \\
			\hline
			\textbf{Right} & 94.75 & 400 & 94.70 & 400 \\
			\hline
			\textbf{Back} & 80.19 & 340 & 80.19 & 340 \\
			\hline
			\textbf{Aggregate} & \textbf{91.36} & \textbf{1500} & \textbf{91.10} & \textbf{1520} \\
			\hline
		\end{tabular}
	}
	\caption{\small{{Results from the classical and annealer-based approximate max set-coverage models with $4 \times 4$ grid, for the free-orientation case.}}}
	\label{tab:results_2_tpo}
	\vspace{-0.3cm}
\end{table}

\begin{figure}
	\begin{subfigure}{0.24\textwidth}
		\hspace{-0.3cm}
		\includegraphics[width=0.19\textheight]{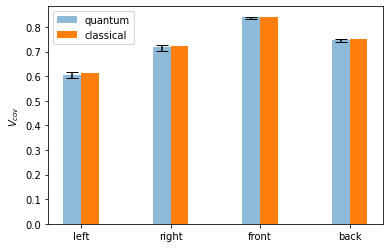}
		\caption{\small{}}
		\label{fig:model2_cov}
	\end{subfigure}
	\begin{subfigure}{0.24\textwidth}
		\includegraphics[width=0.19\textheight]{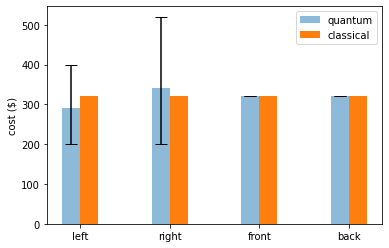}
		\caption{\small{}}
		\label{fig:model2_cost}
	\end{subfigure}
	\caption[ The average and standard deviation of critical parameters ]
	{\small{Coverage and cost of sensors selected by the VQE and classical approaches for all the sides of the downscaled problem.}}
	\label{fig:model2_vqe}
	\vspace{-0.5cm}
\end{figure}

For the gate-model, considering the limited number of qubits available on simulators, a scaled-down version of the problem was considered with only $2 \times 2$ grids, $2$ types of sensors - namely, camera and LiDAR - and orientation of the sensors was fixed to point perpendicularly outwards from the plane. 
The downsized model was optimized using IBM\textsuperscript{\tiny\textregistered} ILOG\textsuperscript{\tiny\textregistered} CPLEX\textsuperscript{\tiny\textregistered}
solver and using a Pennylane-based \cite{pennylane} implementation of VQE, with three layers of the basic entangling layers ansatz (as shown in Figure \ref{fig:basic}). Here too, given the stochastic nature of quantum optimization, $10$ successive quantum  experiments were conducted, the average and min-max range of which has been reported, after the removal of one outlier. A comparison of the classical and quantum results is illustrated in Figure \ref{fig:model2_vqe}, while the coverage and cost of selected configurations with the best objective values are reported in Table \ref{tab:model2_vqe}.

\begin{table}[ht]
	\renewcommand\arraystretch{1.5}
	\centering
	\resizebox{0.48\textwidth}{!}{
		\begin{tabular}{| c | c | c | c | c |}
			\hline
			\multirow{2}{*}{\textbf{Side}} & \multicolumn{2}{| c |}{\textbf{Classical}} & \multicolumn{2}{| c |}{\textbf{Quantum (VQE)}} \\
			\cline{2-1}\cline{3-1}\cline{4-1}\cline{5-1}
			& \textbf{Coverage \%} & \textbf{Cost (\$)} & \textbf{Coverage \%} & \textbf{Cost (\$)} \\
			\hline
			\textbf{Front} & 84.20 & 320 & 84.15 & 320 \\ 
			\hline
			\textbf{Left} & 61.60 & 320 & 61.56 & 320 \\
			\hline
			\textbf{Right} & 72.20 & 320 & 72.19 & 320 \\
			\hline
			\textbf{Back} & 75.10 & 320 & 75.09 & 320 \\
			\hline
			\textbf{Aggregate} & \textbf{86.68} & \textbf{1280} & \textbf{87.77} & \textbf{1280} \\
			\hline
		\end{tabular}
	}
	\caption{\small{{Results for the downsized problem from the classical and VQE-based $(t,p)$ approximate max set-coverage models with $2 \times 2$ grid with only $2$ sensor-types considered.}}}
	\label{tab:model2_vqe}
	\vspace{-0.25cm}
\end{table}

Although the requirement of covering the critical points (with $c_r \geq 0.7$) by at least 2 sensors of different types was not integrated into the models as a constraint, the adherence of the selected sensor-configurations (for the free-orientation case) to this necessity was calculated by collating the sensor-combinations with the best objective values from each side, calculating their FoVs, and checking if the critical points in the RoI lie within the FoVs of:
\begin{enumerate}
	\item two different sensors
	\item two sensors which are of different types.
\end{enumerate}
The extent of satisfaction of this constraint is evaluated by calculating the fraction of relevant points for which the constraint is followed, as presented in Table \ref{tab:constraint}. It is to be noted, however, that a worse adherence implies better solutions found for the aforementioned mathematical models. 

\begin{table}[ht]
	\renewcommand\arraystretch{1.5}
	\centering
	\resizebox{0.48\textwidth}{!}{
		\begin{tabular}{| c | c | c | c |}
			\hline
			\multicolumn{2}{| c |}{\textbf{Approach}} & \textbf{$\textbf{2}$ Sensors} & \makecell{\textbf{$\textbf{2}$ Sensors} \\ \textbf{of Different Types}}\\
			\hline
			\multirow{2}{*}{\textbf{Approach 1}} & \textbf{Quantum (VQE)} & 0.870 & 0.532 \\ 
			\cline{2-1}\cline{3-1}\cline{4-1}&\textbf{Classical} & 0.306 & 0.166 \\ 
			\hline
			\multirow{2}{*}{\textbf{Approach 2}} & \textbf{Quantum (Annealer)} & 0.582 & 0.080 \\ 
			\cline{2-1}\cline{3-1}\cline{4-1}&\textbf{Classical} & 0.334 & 0.078 \\ 
			\hline
			\multirow{2}{*}{\makecell{\textbf{Approach 2} \\ \textbf{(downscaled)}}} & \textbf{Quantum (VQE)} & 0.182 & 0.182 \\ 
			\cline{2-1}\cline{3-1}\cline{4-1}&\textbf{Classical} & 0.171 & 0.171 \\ 
			\hline
		\end{tabular}
	}
	\caption{\small{{Fraction of points (with $c_r \geq 0.7$) in the RoI which are covered by at least two sensors, and two sensors of different types.}}}
	\label{tab:constraint}
	\vspace{-0.5cm}
\end{table}

\section{Conclusion}\label{sec:conclusions}
Two approaches to solve the sensor-placement problem were presented and their classical and quantum implementations were discussed in detail. The results across the two paradigms of computing, and from both the methods show strong conformity. While the first approach uses a lower number of qubits, the appropriate number of sensors to be selected is also an output from the second. The cost-performance of the second approach was noticeably better, for a slight reduction in the coverage. Further, the suggested approximation to the max set-coverage algorithm, which uses a significantly lower qubit-count compared to the traditional Ising formulation, is generic enough to find applications in use cases across various domains. The results obtained from it may then be passed on to a classical evolutionary algorithm, as a warm-starting point, to overcome the suboptimality introduced by the approximation.

\balance

\bibliographystyle{IEEEtran}
\bibliography{SPO.bib}

\end{document}